\documentclass[traditabstract]{aa}
\usepackage{natbib}
\usepackage{epsfig}
\usepackage{txfonts}
\usepackage{graphicx}
\usepackage{float}

\usepackage{color,hyperref}
\definecolor{darkblue}{rgb}{0.0,0.0,0.4}
\hypersetup{colorlinks,breaklinks, linkcolor=darkblue,urlcolor=darkblue, anchorcolor=darkblue,citecolor=darkblue}

\def\m2s2{\hbox{\,m$^{2}$\,s$^{-2}$}} 
\def\Msun{\hbox{$\mathrm{M}_{\odot}~$}}             
\def\Rsun{\hbox{$\mathrm{R}_{\odot}~$}}
\def\Mjup{\hbox{$\mathrm{M}_{\rm Jup}$~}}
\def\Rjup{\hbox{$\mathrm{R}_{\rm Jup}$~}}
\def\Rearth{\hbox{$\mathrm{R}_{\oplus}$~}}
\def\Mearth{\hbox{$\mathrm{M}_{\oplus}$~}}

\def\logg{$\log g$}

\bibpunct{(}{)}{;}{a}{}{,}
\begin{document}
\title{The contribution of secondary eclipses as astrophysical false positives to exoplanet transit surveys}
\subtitle{}

\author{
A. Santerne \inst{1} 
\and F. Fressin \inst{2}
\and R.~F. D\'iaz \inst{3}
\and P. Figueira \inst{1}
\and J.-M. Almenara \inst{3}
\and N.~C. Santos \inst{1,4}
}
\institute{
Centro de Astrof\'isica, Universidade do Porto, Rua das Estrelas, 4150-762 Porto, Portugal
\and Harvard-Smithsonian Center for Astrophysics, Cambridge, MA 02138, USA
\and Aix Marseille Universit\'e, CNRS, LAM (Laboratoire d'Astrophysique de Marseille) UMR 7326, 13388, Marseille, France
\and Departamento de F\'isica e Astronomia, Faculdade de Ci\^encias, Universidade do Porto, Portugal
}
\date{Received: 15 March 2013 ; Accepted: 13 July 2013}

\offprints{Alexandre~Santerne\\
 \email{alexandre.santerne@astro.up.pt}}

\abstract{We investigate the astrophysical false-positive configuration in exoplanet-transit surveys. It involves eclipsing binaries and giant planets that present only a secondary eclipse, as seen from the Earth. To test how an eclipsing binary configuration can mimic a planetary transit, we generated synthetic light curves of three examples of secondary-only eclipsing binary systems that we fit with a circular planetary model. Then, to evaluate its occurrence we modeled a population of binaries in double and triple systems based on binary statistics and occurrence. We find that $0.061\% \pm 0.017\%$ of main-sequence binary stars are secondary-only eclipsing binaries that mimics a planetary transit candidate with a size down to the size of the Earth. We then evaluate the occurrence that an occulting-only giant planet can mimic an Earth-like planet or even a smaller one. We find that $0.009\% \pm 0.002 \%$ of stars harbor a giant planet that only presents the secondary transit. Occulting-only giant planets mimic planets that are smaller than the Earth, and they are in the scope of space missions like \textit{Kepler} and \textit{PLATO}. We estimate that up to $43.1\pm 5.6$ \textit{Kepler} objects of interest can be mimicked by this configuration of false positives, thereby re-evaluating the global false-positive rate of the \textit{Kepler} mission from $9.4 \pm 0.9$\% to $11.3\pm 1.1$\%. We note, however, that this new false-positive scenario occurs at relatively long orbital periods compared with the median period of \textit{Kepler} candidates.
}

\titlerunning{The contribution of secondary eclipses as astrophysical false positives to exoplanet transit surveys}
\authorrunning{A.~Santerne et al.}

\maketitle

\section{Introduction}
\label{intro}
Many astrophysical false positives can mimic an exoplanetary transit. These astrophysical false positives are composed of various configurations of eclipsing binaries (hereafter EB) in which the companion star is gravitationally bound to the target star (EB in double or triple systems) or in the background or foreground within the photometric aperture of the instrument \citep{2009A&A...506..337A}. A transit of a small planet may also be explained by the transit of a larger planet orbiting a background star or a stellar companion of the target star. Recently, \citet{2013arXiv1301.0842F} have investigated the rate of each false positive scenario in the stellar population of the \textit{Kepler} field. By simulating the \textit{Kepler} transit survey using assumptions that are as realistic as possible, they find that the global false-positive rate of \textit{Kepler} is 9.4\%$\pm$ 0.9\%. The authors find that most of false positives involve Neptune-size planets transiting companion stars of the target and mimicking earth-size ones. Their analysis re-evaluates the overall false-positive probability of \textit{Kepler} previously estimated as $\sim$ 5\% by \citet{2011ApJ...738..170M}, which was later found to fall short of the direct estimates by spectroscopic observations \citep{2012A&A...545A..76S}. An accurate estimation of the false positive rate is crucial to determining the occurrence and properties of extrasolar planets from the list of \textit{potential} transiting planets \citep{2012arXiv1202.5852B}. Such studies of planet occurrence and properties in the \textit{Kepler} field have been performed by \citet{2012ApJS..201...15H} but they assume that the rate of false positives could be neglected. These results could be affected if this is not the case, as in fact observed for giant planets, for which both a direct spectroscopic survey \citep{2012A&A...545A..76S} and a dedicated statistical analysis of the detections \citep{2013arXiv1301.0842F}, converge towards a false positive rate between 17.7\% and 35\%. \\

The limitations of current spectrographs do not allow the establishment of the smallest \textit{CoRoT} and \textit{Kepler} planets using the usual radial velocity technique. The planet-validation technique is a new method of establishing the planetary nature of a transiting planet candidate \citep[e.g.][]{2011ApJ...727...24T}. It consists in computing the probability that each false positive scenario has to reproduce the observed data sets. Then, these probabilities are compared with the probability that the transit signal is produced by a bona-fide and undiluted planet. A planet is thus considered as ``validated'' if the planet scenario is significantly the most likely one. An exhaustive set of astrophysical false positive scenarios must be considered in this process to avoid underestimating the false-positive probability of the candidate. \\

The list of currently considered astrophysical false-positive scenarios is explained well in \citet{2012Natur.492...48C} and \citet{2013arXiv1301.0842F}. Briefly, a planetary transit might be mimicked by
\begin{enumerate}
\item an eclipse of an FGK-type main-sequence star by a low-mass star or a brown dwarf with a radius similar to that of Jupiter;
\item an eclipse of a giant star by a main-sequence star;
\item a grazing EB;
\item an eclipse of a binary in the foreground or background, aligned with the target star, as seen from the Earth;
\item an eclipse of a binary bound with the target star;
\item a transit of a planet on a star aligned with the target star, as seen from the Earth, in the foreground or background;
\item a transit of a planet on a star physically bound with the target star,
\item a transiting or occulting white dwarf.
\end{enumerate}

 While the first three scenarios can be easily distinguished by a radial velocity follow-up \citep{2012A&A...545A..76S}, the last ones require rigorous investigations to be statistically rejected. We note that these scenarios may also be constrained thanks to radial velocity diagnosis (Santerne et al., in prep.).\\

We present the false-positive scenario configuration involving secondary-only eclipsing binaries or giant planets. This false-positive scenario is only considered in the false-positive probability estimation of \citet{2011ApJ...738..170M} and planet-validation process described in \citet{2012ApJ...761....6M} and \citet{2012ApJ...761..163D} and is not considered in \textit{Kepler} planet validation and in the false-positive probability estimation of \citet{2013arXiv1301.0842F}. We first present the configuration secondary-only eclipsing binary in Section \ref{newFP} and evaluate its occurrence by simulating a population of eclipsing binaries in Section \ref{occurrenceEB}. We then consider the same configuration in the case of a giant planet and evaluate its occurrence in Section \ref{occurrencePL}. Finally, we discuss its impact on exoplanet-transit surveys like \textit{CoRoT} and \textit{Kepler} in Section \ref{discussion}.

\section{Secondary-only eclipsing binaries as false positive scenario}
\label{newFP}

\subsection{Conditions for secondary-only eclipsing binaries}

\begin{figure}[]
\begin{center}
\includegraphics[width=\columnwidth]{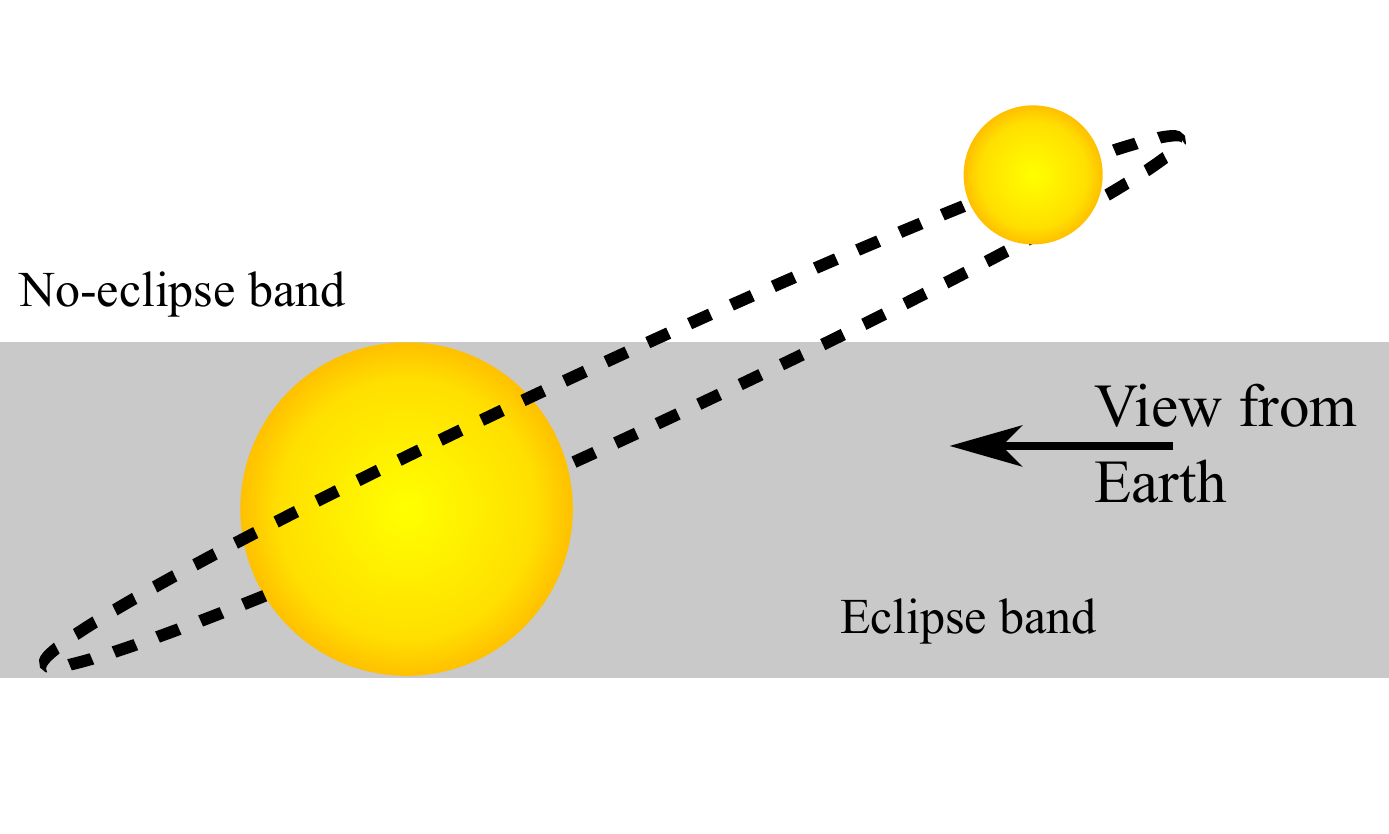}
\caption{Sketch of the configuration where a binary only shows a secondary eclipse.}
\label{SecondaryEclipse}
\end{center}
\end{figure}

In all cases where the planet-validation technique was used to establish the planetary nature \citep[e.g.,][]{2012ApJ...745...81F, 2012Natur.482..195F}, only the primary eclipse -- i.e., when the smaller object passes in front of the larger object -- of EB were considered as the cause of the observed transit. In some geometrical configurations, an EB might \textit{not} present a primary eclipse, but \textit{only} a secondary eclipse -- i.e. when the smaller object is eclipsed by the larger one (see Fig. \ref{SecondaryEclipse}). This configuration was accounted for only in the false-positive studies of \citet{2011ApJ...738..170M}, \citet{2012ApJ...761....6M}, and \citet{2012ApJ...761..163D}, but they do not provide any detailed statistics or occurrence rate. For the first time, \citet{2012A&A...545A..76S} characterized two eclipsing binaries among the \textit{Kepler} transiting exoplanets candidates (namely KOI-419 and KOI-698) that only present their secondary eclipse, as seen from the Earth. A secondary eclipse of EBs can be as shallow as to mimic a transit event of a planet \citep[see also Fig. \ref{Posterior}]{2012A&A...545A..76S}. This configuration is only possible for orbits for which the impact parameters ($b_{prim}$ and $b_{sec}$) satisfy \citep{2010arXiv1001.2010W}
\begin{eqnarray}
b_{prim} &=& \frac{a}{R_{1}}\cos(i) \left(\frac{1-e^{2}}{1+e\sin\omega}\right) > 1 + \frac{R_{2}}{R_{1}},\label{eq1}\\
b_{sec} &=&\frac{a}{R_{1}}\cos(i) \left(\frac{1-e^{2}}{1-e\sin\omega}\right) < 1 + \frac{R_{2}}{R_{1}},\label{eq2}
\end{eqnarray}
where $a$ is the semi-major axis, $R_{1}$ and $R_{2}$ the radii of the primary and secondary stars (respectively), $i$ the orbital inclination, $e$ the orbital eccentricity, and $\omega$ the argument of periastron. To present only a secondary eclipse, the orbit should have nonzero eccentricity, with an argument of periastron $\omega \in \left[180^{\circ};360^{\circ}\right]$ and an orbital inclination different from $90^{\circ}$. The eccentricity of the system is the main parameter driving this false-positive configuration.\\

\subsection{Eccentricity of short-period binaries}
\label{ESPB}

\begin{figure}[]
\begin{center}
\includegraphics[width=\columnwidth]{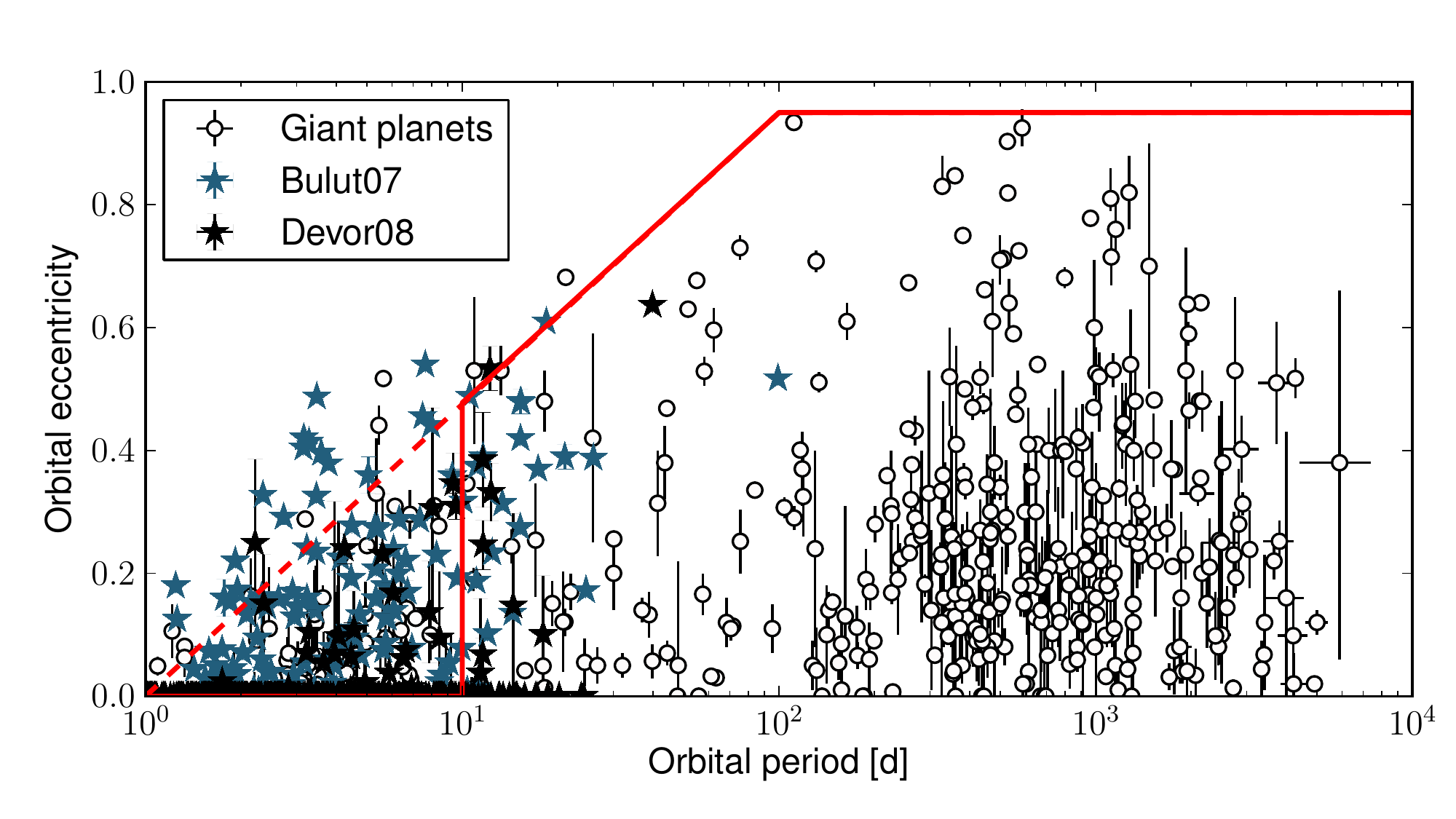}
\caption{Period -- Eccentricity diagram of giant planets discovered by radial velocity (open circles) and eclipsing binaries from TrES \citep[black stars,][]{2008AJ....135..850D} \and from the catalog of eccentric eclipsing binaires \citep[blue stars,][]{2007MNRAS.378..179B}. The solid red line displays the assumed upper-envelope of the eccentricity of binaries, and the dashed red line, the upper-envelope of eccentricity of giant planets.}
\label{PerEccPrior}
\end{center}
\end{figure}

It has been reported by \citet{1991A&A...248..485D}, \citet{2003A&A...397..159H}, and \citet{2010ApJS..190....1R} that main-sequence eccentric binaries do not have orbital periods of less than about ten days. This circularization period is compatible with the theory of the tidal evolution of close-in binaries \citep{1977A&A....57..383Z, 1989A&A...220..112Z}. On the other hand, numerous detached-EB from the \textit{Kepler} catalog \citep{2011AJ....142..160S} are eccentric with orbital periods down to about four days (see for example KIC4947726 and KIC4753561 in the \textit{Kepler} EB catalog). A thorough analysis of EB in this catalog can reveal their distribution of eccentricities, which is beyond the scope of this paper. This has been performed based on the Trans-Atlantic Exoplanet Surveys \citep{2008AJ....135..850D} that confirm the nonzero eccentricity of some eclipsing binaries with orbital period shorter than ten days (as displayed in Fig. \ref{PerEccPrior}). However, significant eccentricity values in such short-period binary system have been reported for only a very few cases. Indeed, among the 827 EBs (respectively 725) reported by \citet{2008AJ....135..850D} with orbital period of less than ten days (respectively, five days), only five systems (respectively three systems) present a significant eccentricity ($>3-\sigma$). This corresponds to $0.6\pm0.3$\% of their total sample of EBs (respectively, $0.4\pm0.2$\%). Moreover, we cannot exclude that these systems are actually orbiting in a wider stellar system in which the Kozai mechanism \citep{1962AJ.....67..591K} can counterbalance the tidal circularization.\\

On the other hand, \citet{2007MNRAS.378..179B} compiled a catalog of 124 eccentric eclipsing binaries and 150 candidates reported in the \textit{HIPPARCOS} catalog \citep{1997A&A...323L..49P}, the atlas of (O-C) diagram of \citet{2001aocd.book.....K} and the ninth catalog of spectroscopic binary orbits \citep[$S_{B^{9}}$,][]{2004A&A...424..727P}. The majority of these binaries have an orbital period of less than ten days (see Fig. \ref{PerEccPrior}). From the $S_{B^{9}}$ catalog, 16\% of spectroscopic binaries present an orbital period of less than ten days and a significant eccentricity (assumed $> 0.1$) among 1515 systems of various spectral types and luminosity classes. We supposed that these eccentric short-period binaries might be dominated by systems with a massive primary with no convective zone to dissipate tidal forces. Those systems might also be younger than the circularization timescale. \citet{2005ApJ...629..507A} collected nearly 400 systems from $S_{B^{9}}$ having a B0 -- F0 V -- IV primary. Sixty percent (respectively, 23\%) of such systems have an eccentricity greater than zero (respectively, greater than 0.1). Binaries with massive primary are thus less circularized than the population of binaries reported in $S_{B^{9}}$. Since transit surveys focus mainly on FGK(M) dwarfs, we therefore assume that this configuration of EB can occur at any orbital period, but is relatively rare for orbital periods less than ten days.\\

\subsection{Shape of secondary-only eclipse}
\label{SSOE}

\begin{figure}[]
\begin{center}
\includegraphics[width=\columnwidth]{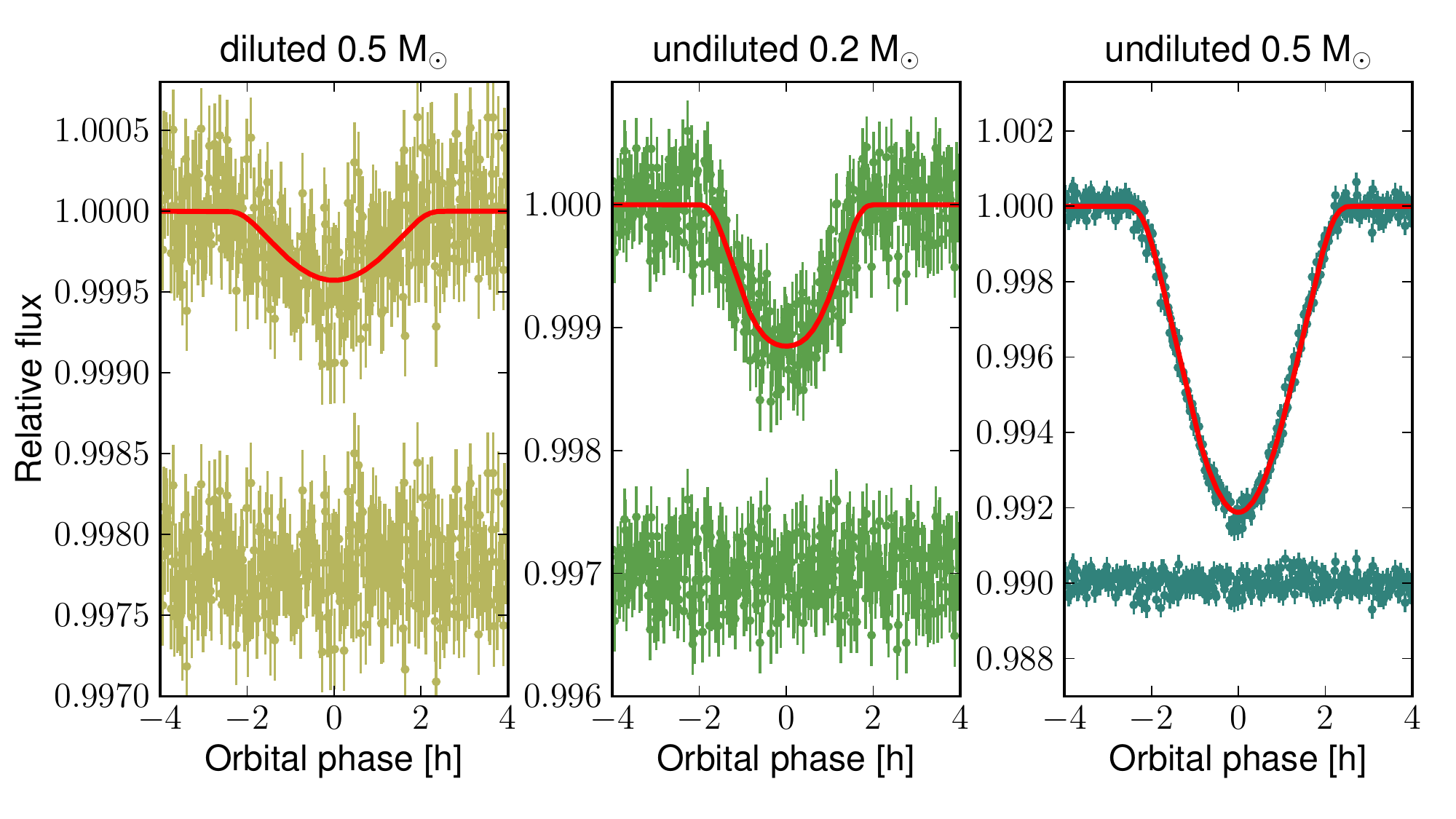}
\includegraphics[width=\columnwidth]{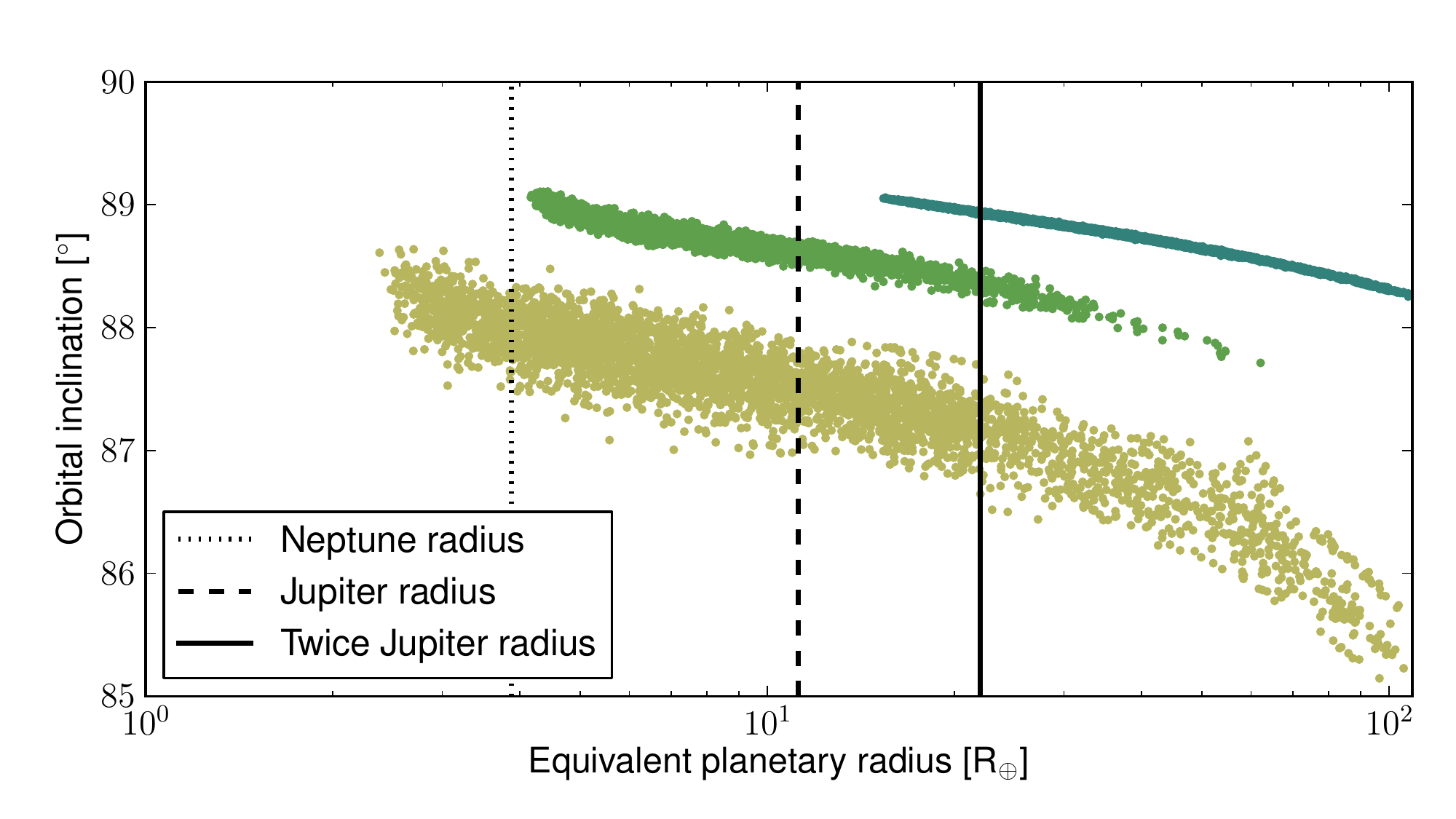}
\caption{(Top) Synthetic \textit{Kepler} light curve of secondary-only EB (as described in the text) with their best planetary model and residuals. From left to right, light curve are generated assuming the 1\Msun -- 0.5\Msun secondary-only EB diluted by a gravitationally bound 1\Msun star, the undiluted 1\Msun -- 0.2\Msun secondary-only EB and the undiluted 1\Msun -- 0.5\Msun secondary-only EB. (Bottom) Posterior distribution from the MCMC analysis of the aforementioned secondary-only EBs, displaying the correlation between the orbital inclination and the planetary radius (assuming a 1-\Rsun host). Colors and ranking from left to right are the same as for the upper plot. Vertical lines indicate the radius of Neptune (dotted line), the radius of Jupiter (dashed line), and twice the radius of Jupiter (solid line).}
\label{SOtest}
\end{center}
\end{figure}

Secondary-only eclipses can easily mimic the depth of a planetary transit as shown with KOI-419 and KOI-698 \citep{2012A&A...545A..76S}. Their transit shapes are either ``V-shaped'' (grazing eclipses, see Fig. \ref{SOtest}) or do not have limb-darkening effects (total eclipses). In all cases, the transit-like event has a relatively short duration (see Fig. \ref{Posterior}). However, several planets have already been reported with short and grazing transits, e.g. CoRoT-10~b \citep{2010arXiv1006.2949B}, showing that V-shape transit can be compatible with planets. \\

To further test how a secondary-only EB can mimic a planetary transit, we generated synthetic light curves of three binary systems with the \texttt{PASTIS} code \citep{diazetal13} for the \textit{Kepler} bandpass, using the \texttt{Ebop} code \citep[and references therein]{2008MNRAS.386.1644S} and stellar atmosphere models from the PHOENIX/BT-Settl library \citep{2012IAUS..282..235A}. For the three systems, we assumed an orbital period of $\sim$63.1 days, inclination of 89$^{\circ}$, eccentricity of 0.3, and argument of periastron of 270$^\circ$. These values are intended to represent the median values of the distribution of secondary-only eclipsing binaries (see Fig. \ref{Posterior} and section \ref{DiscutDistribSOEB}). For the first system, we assumed two stars with masses of 1\Msun and 0.5\Msun that produce a secondary-only eclipse at the level of $\sim$~1\%. For the second system, we assumed two stars with masses of 1\Msun and 0.2\Msun that produce a secondary-only eclipse at the level of $\sim$~0.1\%. Finally, the third system is composed of two stars of 1\Msun for which one host a secondary-only eclipsing companion of 0.5\Msun. This system produces a diluted secondary-only eclipse at the level of $\sim$~500~ppm. We assumed only white noise with an amplitude of 250~ppm, which is the typical precision of the \textit{Kepler} machine for a magnitude K$_{p}=15$ target. Synthetic light curves have an integrated sampling of 30 minutes to reproduce the \textit{Kepler} long-cadence data and a timescale of 3.5 years. Synthetic light curves are displayed in Fig. \ref{SOtest}. \\

We fit the three generated light curves with a planetary scenario using the MCMC algorithm of the \texttt{PASTIS} code \citep{diazetal13} that includes a principal component analysis decomposition to better explore the correlated parameter space. We fixed the eccentricity to zero and the limb darkening values to the solar ones from the table of \citet{2012A&A...546A..14C}. Only the orbital period, transit epoch, system scale $(a/R_{\star})$, radius ratio, orbital inclination, and out-of-transit flux were left as free parameters in the MCMC analysis. In Fig. \ref{SOtest} (bottom) we represent the correlation between the orbital inclination and the measured planetary radius from the posterior distribution of the MCMC analysis, assuming a 1-\Rsun stellar host. The MCMC fit converged toward a stellar density lower than the Sun, which is still compatible with the values observed for transiting planets \citep{2011ApJ...726..112T}. This lower stellar density can also be explained by an eccentric planet \citep{2012ApJ...756..122D, 2012ApJ...761..163D}. The three synthetic light curves of secondary-only EB are thus compatible with a planet with a radius below twice the one of Jupiter. We note that above the commonly-used 2-\Rjup limit for giant planets, objects are most likely of stellar origin and cannot be fit assuming a non-self-emitting object as in case of a planet. The MCMC analysis thus explored unphysical models above the 2R$_\mathrm{jup}$-limit. The goal here was to convince skeptical readers that the degeneracy between radius ratio and orbital inclination in case of a V-shape transit allow grazing eclipsing binary to mimic planets. The best planetary models that satisfy $r_{p} < 2$\Rjup are displayed in Fig. \ref{SOtest} (top), with their residuals.\\

From the transit parameters of \citet{2012arXiv1202.5852B}, we find that about 60\% (respectively, 92\%) of the \textit{Kepler} objects of interest (KOIs) have an impact parameter greater than one within 1-$\sigma$ (respectively,  3-$\sigma$). Since the majority of the KOIs are compatible with a grazing transit or V-shape transit within 3-$\sigma$, we therefore do not consider the shape of a transit event as a systematic indication of false positive. We stress that \textit{Kepler} has a photometric precision high enough to \textit{detect} the transit of very small candidates. Unfortunately, the signal-to-noise reached by \textit{Kepler} is too low for the majority of the KOIs to \textit{determine the shape} of the transit, especially if the candidate is small, the orbital period is long, the host star is active and/or the host star is faint. This does not mean that the majority of the KOIs are actually V-shaped transit.\\

Beaming, ellipsoidal, and reflexion effects \citep{2010A&A...521L..59M} might be used to identifying eclipsing binary that would produce out-of-transit variations. However, as shown by \citet{2011AJ....142..195S}, the amplitude of these effects drastically decrease with the orbital period: the beaming effect decreases as $a^{-1/2}$, the ellipsoidal effect decreases as $a^{-3}$, and the reflexion effect as $a^{-2}$ (where $a$ is the orbital semi-major axis). Recently, \citet{2012ApJ...746..185F} characterized seven non-eclipsing binaries thanks to these effects, but none of them present an orbital period longer than ten days. Assuming the same systems but with a orbital semi-major axis ten times larger, the beaming, reflexion, and ellipsoidal effects would present an amplitude at the level of the 100~ppm, 2~ppm, 1~ppm (respectively), or below. We therefore consider that these effects can only marginally be used to distinguish secondary-only eclipsing binaries from out-of-transit variations, especially if the primary star is active.

\section{Estimation of the occurrence rate of secondary-only eclipsing binaries}
\label{occurrenceEB}

\subsection{Modeling the population of multiple stellar systems}

To evaluate the occurrence of this EB configuration, we performed a Monte Carlo simulation by considering $10^{6}$ binaries, hierarchical triple systems (where the inner-binary is composed of the two lowest mass star), and nonhierarchical triple systems (where the brightest star is member of the inner-binary) based on the statistics reported in \citet{2010ApJS..190....1R}. We thus assumed the following distributions:
\begin{itemize}
\item log-normal distribution of orbital period ($P$) either for the binary, inner-, and outer-binary of a triple system, centered at $\log_{10}P = 5.03$ with a width of $\sigma_{\log_{10}P} = 2.28$ \citep{2010ApJS..190....1R} ; 
\item distribution of primary mass of the \textit{Kepler} targets, according to the \textit{Kepler} Input Catalog \citep{2011AJ....142..112B}, which is representative of a population of solar-type main-sequence stars outside of the galactic plane. After rejecting targets with \logg~$< 3.6$, these stars range in mass from $\sim$~0.3~\Msun to $\sim$~2.1~\Msun, with a sharp maximum close to 1~\Msun;
\item mass-ratio distributions following figure 16 of \citet{2010ApJS..190....1R}. We do not consider brown dwarfs owing to the limitation of our isochrones and their relatively rare occurrence ;
\item sine distribution of orbital inclination \citep{2012A&A...541A.139F} ;
\item uniform distribution of $\omega$ between $0^{\circ}$ and $360^{\circ}$ ;
\item a circular orbit for systems with orbital periods less than ten days; a uniform distribution between a zero eccentricity and a linear upper-envelope ranging in eccentricity in $\left[0.5; 0.95\right]$ and in period in $\left[10; 100\right]$ days (see Fig. \ref{PerEccPrior}). For periods longer than 100 days, we assumed a uniform distribution in the range $\left[0; 0.95\right]$ (see Fig. \ref{PerEccPrior}).
\end{itemize}

We limit the eccentricities to a reasonable maximum value ($e<0.95$) to avoid overestimating the fraction of secondary-only EB. The stability of triple systems was tested using eq. 16 of \citet{2013arXiv1302.0563R} and unstable systems were not allowed. Stellar radius were estimated using an isochrone of 1~Gyr from the Dartmouth Stellar Evolution Database \citep{2008ApJS..178...89D}. For each binary that satisfies equations \ref{eq1} and \ref{eq2}, we computed the secondary-eclipse depth with the \texttt{PASTIS} code \citep{diazetal13} for the \textit{Kepler} bandpass as described in the previous section.

\subsection{Occurrence and comparison with other result}

The fractions of EBs found are reported in Table \ref{table} for each type (secondary-only, primary-only, and both primary and secondary), and their distributions are displayed in figures \ref{Posterior} and discussed in section \ref{DiscutDistribSOEB}. 
To compute the occurrence of each scenario, we multiplied the fraction (among our $10^{6}$ simulated cases) of such a scenario with the overall likelihood that a given star follow this scenario \citep{2010ApJS..190....1R, 2013arXiv1303.3028D}. Following \citet{2013arXiv1301.0842F}, we corrected the overall likelihood by the mass of the primary to reflect that massive stars are more common in multiple system than low-mass stars \citep[see Fig. 12 in][]{2010ApJS..190....1R}.
We find that $0.043\% \pm 0.004\%$ of stars are secondary-only EBs and $0.030\% \pm 0.010\%$ of stars are secondary-only EBs in triple systems. Considering only those that present a depth shallower than 3\%, we find an occurrence rate of secondary-only EB of $0.061\% \pm 0.017\%$ for primary of spectral type FGK IV--V. Uncertainties were estimated by considering the uncertainty of our simulation (assuming a Poisson noise) and the uncertainty on the occurrence of binaries from \citet{2010ApJS..190....1R}. We also accounted for the uncertainty on the occurrence of the different hierarchies in triple systems \citep{2010ApJS..190....1R}. Allowing short-period EB (less than ten days) to be eccentric in our simulation, using the same prior distribution as for binaries with period longer than ten days, we found an upper occurrence of secondary-only EB of $0.082 \pmÊ0.025$\%. Thus, considering the eccentricity of short-period binaries or not does not change the occurrence rate within 1-$\sigma$.\\

\citet{2012A&A...545A..76S} characterized two secondary-only eclipsing binaries using radial velocity observations with the SOPHIE spectrograph \citep{2009A&A...505..853B}. These two false positives were observed in a selection of 46 close-in giant candidates with orbital periods of less than 25 days, transit depths deeper than 0.4\% , and a host star brighter than K$_{p}=14.7$. From the MAST archive, we found 63542 dwarfs (with \logg $> 4.0$) brighter than K$_{p}=14.7$ observed by \textit{Kepler} since 2009. From our simulation, selecting only the secondary-only EB with orbital periods of less than 25 days and with eclipse depths between 0.4\% and 3\%, we expect $4.3 \pmÊ0.5$ false positives in the \citet{2012A&A...545A..76S} sample. This discrepancy might be explained either by some secondary-only EB in the eight candidates flagged with a vetting of four by \citet{2012ApJ...745..120B} that were not observed with SOPHIE. This discrepancy might also be explained by some secondary-only EB that would have been identified as false positive prior to ground-based observations in the vetting process performed by the \textit{Kepler} team. This discrepancy, based on small number statistics, might also reveal the overestimation of this false-positive scenario in our simulation.\\

\citet{2011AJ....142..160S} report an occurrence rate of detached-EB of 0.79\% in the \textit{Kepler} EB catalog which is significantly higher than our estimation listed in Table \ref{table}. First of all, we believe that the occurrence of EB in the \textit{Kepler} catalog is slightly over estimated since it is composed of several confirmed planets, such as KIC9818381 also known as KOI-135~b / Kepler-43~b \citep{bonomo2012}, KIC5728139 - KOI-206~b \citep[Almenara et al. in prep.;][]{2012A&A...545A..76S}, or Kepler-76~b \citep{2013arXiv1304.6841F}. Then, the \textit{Kepler} EB catalog is composed of EB that do not involve the target star. Accounting for all EB from our simulation that present at least a primary or secondary eclipse in our model with a depth greater than 3\%, we found an occurrence of EB of 0.53\% $\pm$ 0.14\% that is compatible at 1.8-$\sigma$ with the occurrence of EB found in the \textit{Kepler} field.\\

\begin{table*}[]
\renewcommand{\footnoterule}{}                          
\begin{minipage}[c]{\textwidth} 
\caption{1: Fraction of eclipsing binaries in double and triple systems and transiting giant planets, among $10^{6}$ simulated systems, which present either the primary or secondary or both primary and secondary eclipse(s), as seen from the Earth. 2: Global occurrence rate for these systems, as reported in the literature. 3: Occurrence rate (fraction $\times$ overall likelihood $\times$ spectral type correction) of secondary-only, primary-only, and both primary and secondary eclipse(s) for the different configurations of system. 4: Number of KOIs that might be mimicked by a secondary-only system according to the \textit{Kepler}-capability detection model of \citet{2013arXiv1301.0842F}.}
\begin{center}
\begin{tabular}{lcccc}
 \hline
1. & Binary system & Hierarchical triple & Nonhierarchical triple & Giant planet\\
 Fraction of secondary-only & $0.126\% \pm 0.004\%$ & $0.227\% \pm 0.005\%$ & $0.391\% \pm 0.006\%$ & $0.091\% \pm 0.003\%$ \\
 Fraction of primary-only & $0.129\% \pm 0.004\%$ & $0.228\% \pm 0.005\%$ & $0.394\% \pm 0.006\%$ & $0.482\% \pm 0.007\%$ \\
 Fraction of primary and secondary & $0.633\% \pm 0.008\%$ & $2.153\% \pm 0.015\%$ & $3.632\% \pm 0.019\%$ & $1.703\% \pm 0.013\%$ \\
Total fraction of primary and/or secondary & $0.888\% \pm 0.016\%$ & $2.608\% \pm 0.025\%$ & $4.417\% \pm 0.031\%$ & $2.276\% \pm 0.023\%$ \\
 \hline
& & & & \\
& & & & \\
 \hline 
2. & Binary system & Hierarchical triple & Nonhierarchical triple & Giant planet\\
 Overall likelihood & 33\%$\pm$2\%$^{\dag}$ & 11\%$\pm$2\%$\times$ 76\%$^{\dag}$ & 11\%$\pm$2\%$\times$ 24\%$^{\dag}$ & 9.7\% $\pm$ 1.3\%$^{\ddag}$\\
 \hline
& & & & \\
& & & & \\
 \hline 
3. & Binary system & Hierarchical triple & Nonhierarchical triple & Giant planet\\
 Occurrence of secondary-only & $0.043\% \pm 0.004\%$ & $0.019\% \pm 0.007\%$ & $0.011\% \pm 0.003\%$ & $0.009\% \pm 0.001\%$ \\
 Occurrence of primary-only & $0.044\% \pm 0.004\%$ & $0.020\% \pm 0.007\%$ & $0.011\% \pm 0.003\%$ & $0.047\% \pm 0.007\%$ \\
 Occurrence of primary and secondary & $0.215\% \pm 0.016\%$ & $0.184\% \pm 0.062\%$ & $0.099\% \pm 0.025\%$ & $0.165\% \pm 0.023\%$ \\
Total occurrence of primary and/or secondary & $0.302\% \pm 0.022\%$ & $0.223\% \pm 0.075\%$ & $0.120\% \pm 0.031\%$ & $0.221\% \pm 0.031\%$ \\
 \hline
& & & & \\
& & & & \\
 \hline 
4. & Binary system & Hierarchical triple & Nonhierarchical triple & Giant planet\\
Number of mimicked KOIs & $22.6 \pm 4.4$ & $10.0 \pm 2.7$ & $10.1 \pm 2.7$ & $0.4  \pm 0.4$\\
 \hline
\end{tabular}
\end{center}
\label{table}
\vspace{-0.5cm}
\footnotetext{$^{\dag}$ As reported by \citet{2010ApJS..190....1R}; $^{\ddag}$ As reported by \citet{2011arXiv1109.2497M} for giant planets at any orbital period ($m_{p}\sin i > 100\, \Mearth$)}
\end{minipage}
\end{table*}

We tested the dependence of our results on the assumed prior distributions. We expect the period distribution to have a significant impact on the resulting occurrence, especially for the inner binary of the triple system. All inner binary of triple systems reported by \citet{2010ApJS..190....1R} present an orbital period of less than 100 days. Limiting the periods of such binaries to 100 days increases the fractions and occurrences of the hierarchical and nonhierarchical triple systems reported in Table \ref{table} by factors of 1.43 and 1.46 (respectively). In that case, our value of the occurrence of EB would be in better agreement with the value from \citet{2011AJ....142..160S}. The primary-mass distribution is expected to significantly affect the results. Assuming the population of F -- K dwarfs with $11 < m_{R} < 16$ located in each of the \textit{CoRoT} eyes \citep{2006ESASP1306...19B}, as simulated with the Besan\c con galactic model \citep{2003A&A...409..523R}, we found a lower occurrence of $20 \pmÊ9\%$ and $16\pm6$\% for the center and anticenter fields, respectively. This might be explained by the fact that dwarfs in the \textit{CoRoT} eyes are on average smaller than the \textit{selected} \textit{Kepler} targets, according to the Besan\c con galactic model. Thus, both their eclipse probability and binary occurrence rates are lower. Finally, assuming the eccentricity distribution reported by \citet{2010ApJS..190....1R} for binaries with periods below 1000 days, within our envelope displayed in  Fig. \ref{PerEccPrior}, we find a lower occurrence of secondary-only EB of $0.039\pm0.012$ \%, at 1-$\sigma$ from the occurrence rate found with a uniform distribution of eccentricity.

\subsection{Distributions of secondary-only eclipsing binaries}

\label{DiscutDistribSOEB}

\begin{figure*}[]
\begin{center}
\begin{tabular}{cc}
\includegraphics[width=\columnwidth]{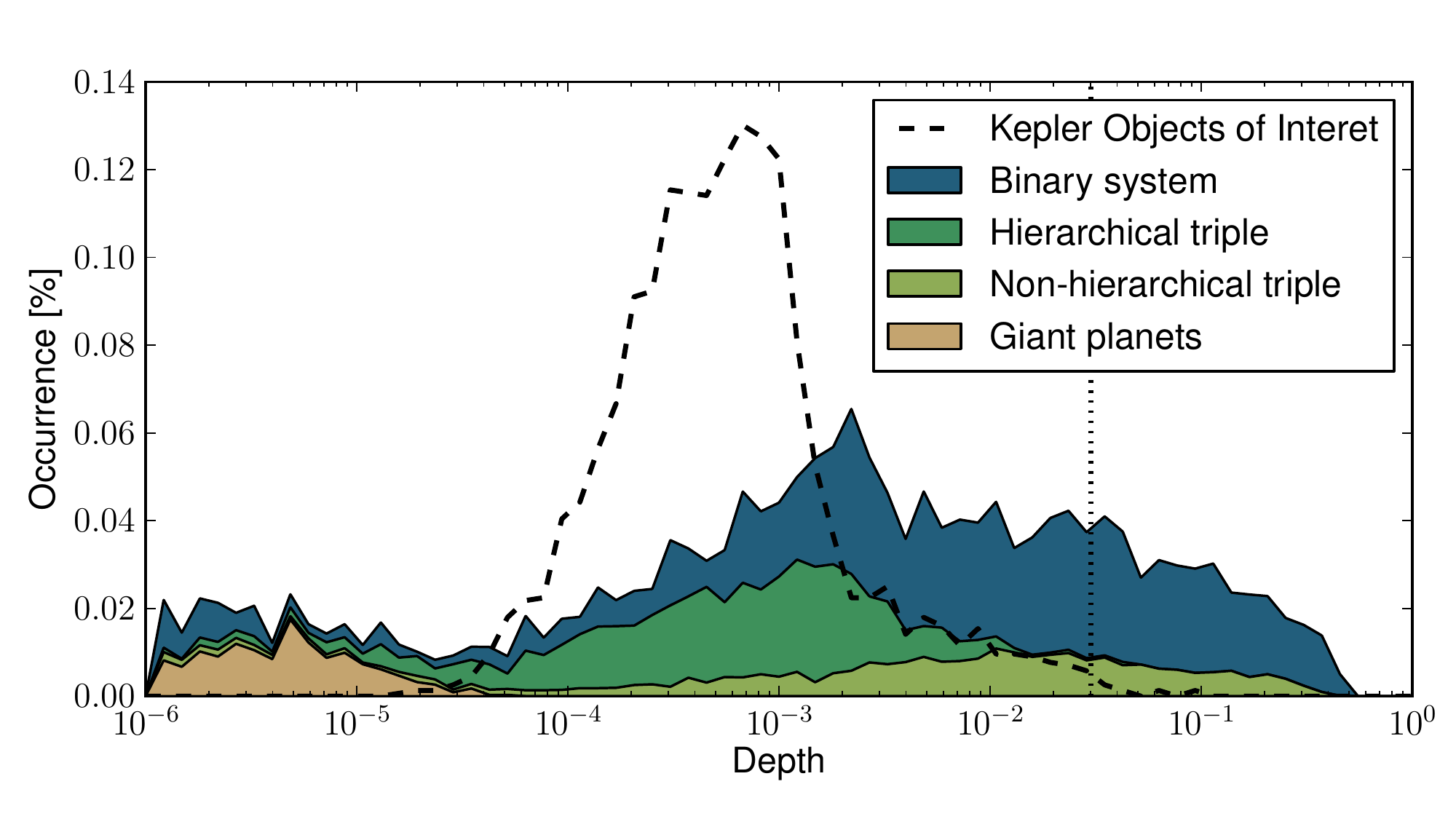} & \includegraphics[width=\columnwidth]{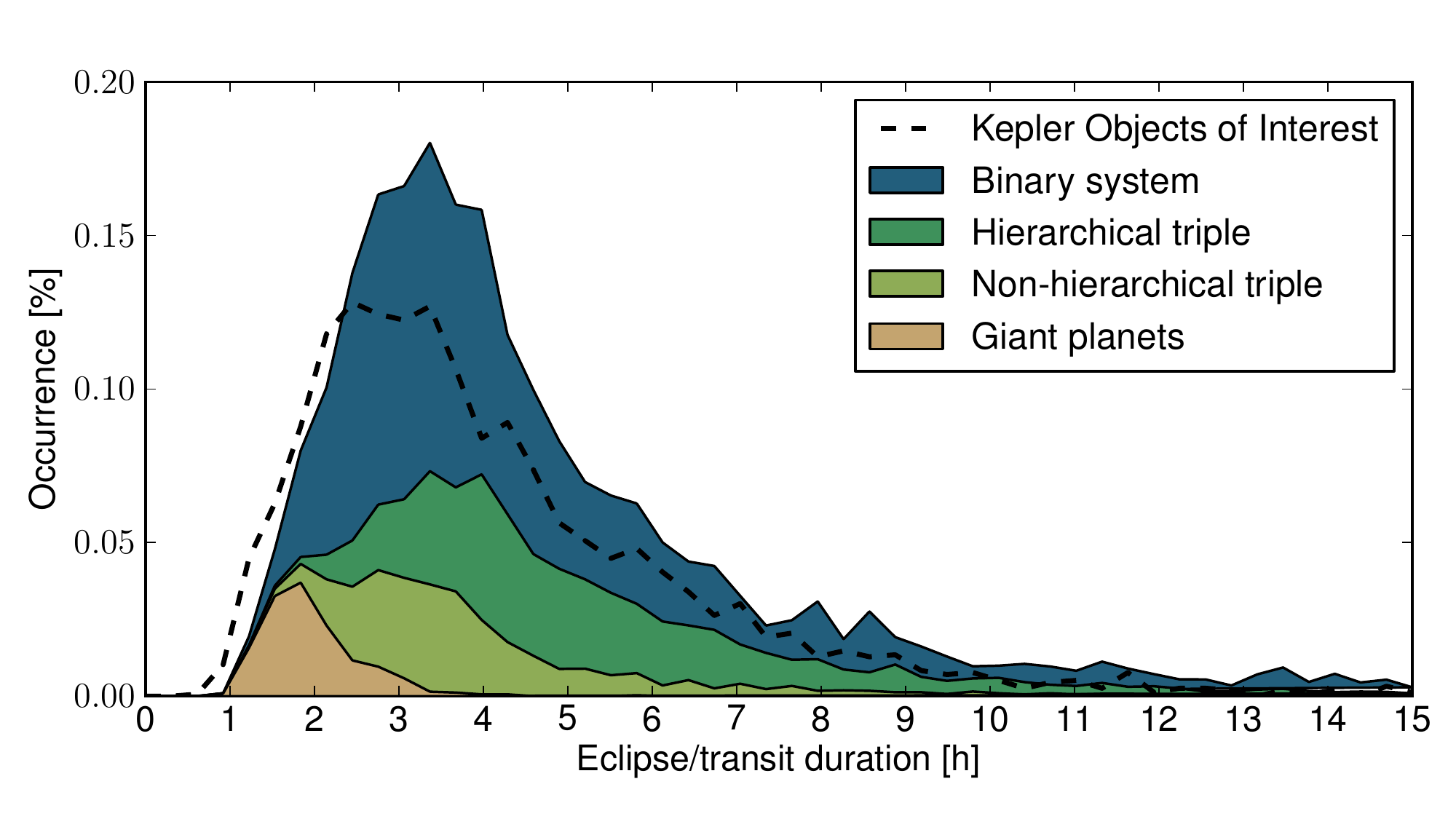} \\
\includegraphics[width=\columnwidth]{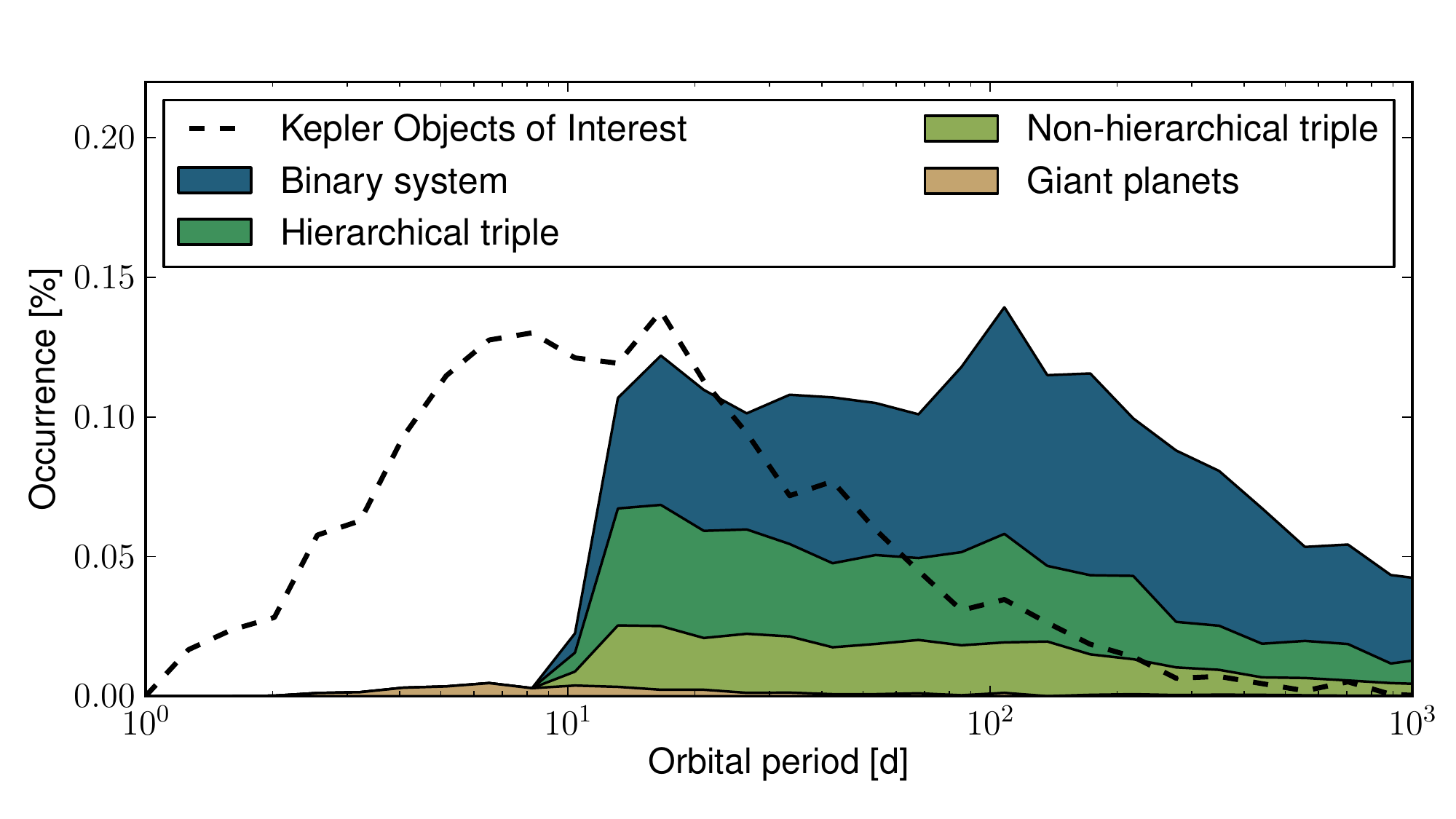} & \includegraphics[width=\columnwidth]{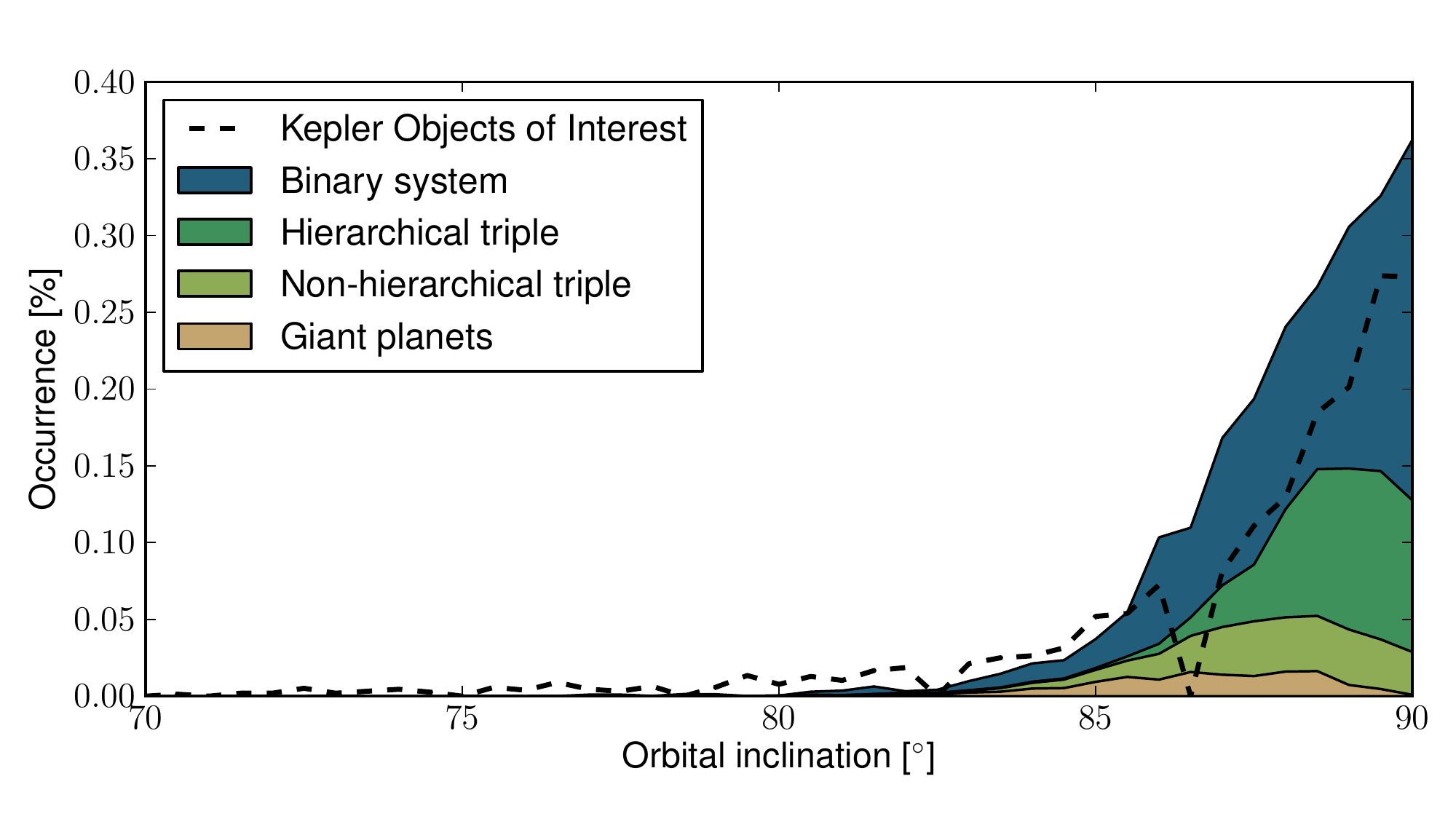}\\
\includegraphics[width=\columnwidth]{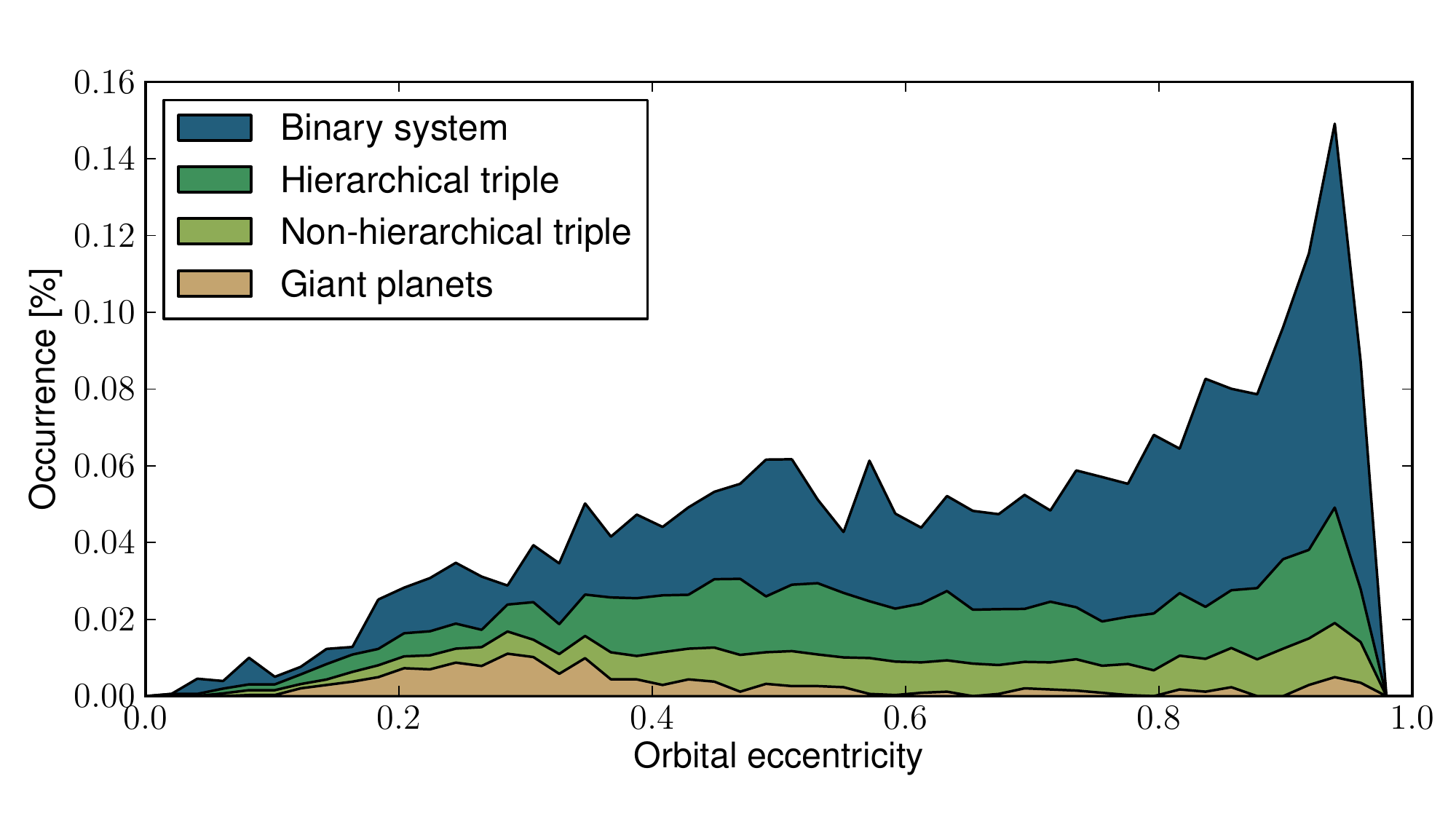} & \includegraphics[width=\columnwidth]{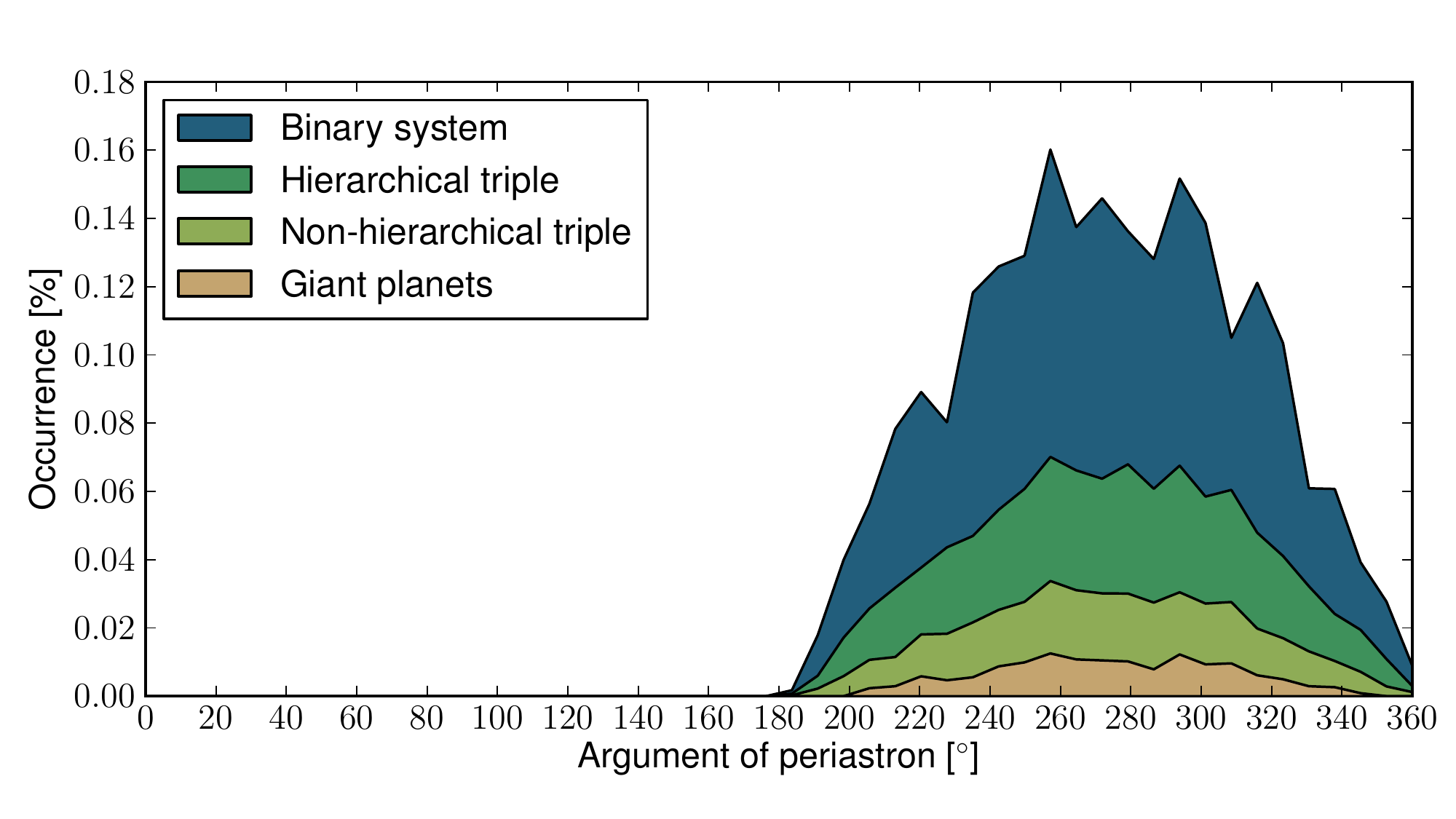}
\end{tabular}
\caption{Stacked distributions (magnified by 30) of the secondary-only EB and occulting-only giant planets for their eclipse/transit depth (upper-left plot), eclipse/transit duration (upper-right plot), orbital period (middle-left plot), inclination (middle-right plot), eccentricity (lower-left plot), and argument of periastron (lower-right plot). The distribution have been normalized to represent the relative occurrence as computed in Table \ref{table}. Corresponding distributions of \textit{Kepler} objects of interest \citep[from][dashed black line]{2012arXiv1202.5852B} are also displayed, when known. The vertical dotted line in the upper-left plot represent the commonly-used 3\% upper limit in depth of planetary transit candidates.}
\label{Posterior}
\end{center}
\end{figure*}

Figure \ref{Posterior} displays the stacked distribution of depth and transit duration of the simulated binaries that present a secondary-only eclipse. The eclipse depth were estimated as the minimum point of the synthetic light curve. This was only possible due to the good sampling of the synthetic light curve (samping of $10^{-4}$ in phase). Eclipse durations were estimated using the following equation, from \citet{2010arXiv1001.2010W}:

\begin{equation}
T_{14} = \frac{P}{\pi}\sin^{-1}\left[\frac{R_{1}}{a}\frac{\sqrt{\left(1+\frac{R_{2}}{R_{1}}\right)^{2} - b_{sec}^{2}}}{\sin i}\right]\frac{\sqrt{1-e^{2}}}{1-e\sin\omega}.
\end{equation}
The orbital period, inclination, eccentricity, and argument of periastron distribution of secondary-only eclipsing binaries are displayed in Figure \ref{Posterior}. We note that secondary-only EB can mimic the transit depth of planet-candidate for the whole range of radius detectable by \textit{Kepler}. However, secondary-only EB in binary system or nonhierarchical systems more likely mimic giant planets, while secondary-only EBs in hierarchical triple mimic planet candidates with the size of Neptune.\\

Secondary-only EB present eclipse duration with a median of 3.95 hours, while the median value for all the KOIs is 3.41 hours. Nevertheless, as shown in Fig. \ref{Posterior} and discussed in section~\ref{ESPB}, secondary-only EB have an orbital period greater than about ten days for the vast majority of cases. The median value of transit duration among the KOIs with period of more than ten days is 4.65 hours, which is slightly longer than the one estimated for the secondary-only EB. As discussed in section \ref{SSOE}, the shorter duration of secondary-only eclipse might be interpreted as an eccentric transiting planet.\\

Due to the effect of tidal circularization for binaries with orbital period of fewer than ten days, we did not include eccentric binaries with such short orbital periods in our simulation. Secondary-only EB thus present a orbital period longer than ten days, with a median of $\sim$ 134 days for a binary system, and $\sim$ 63 days and 61 days for hierarchical and nonhierarchical triples, respectively. Accounting for the respective occurrence of the different configuration of multiple system, secondary-only EB have an orbital period with a median of 116 days, while the median period of the KOIs is 11 days. This new configuration of false positive is thus expected to be more frequent for long-period candidates. The distribution of orbital inclination deduced from our simulation is not obviously different from the one observed in the KOIs ones, which are dominated by the geometrical transit/eclipse probability.\\

The eccentricity of secondary-only EB is the most important orbital parameter for the configuration of false positives presented in this paper. The eccentricity of secondary-only EB have relative high-eccentricity, with a median of $\sim$ 0.7 (0.73 for those in binary system and 0.61 for those in triple system). As expected by equations \ref{eq1} and \ref{eq2}, we find a posterior distribution of argument of periastron in the range [180$^{\circ}$;360$^{\circ}$], centered on 270$^{\circ}$. These values of $\omega$ are the only ones that allow only the secondary eclipse to be seen.

\section{Occulting-only giant planets as false positive scenario}
\label{occurrencePL}

\subsection{Modeling the population of giant planets}

We now consider the occultation of a giant planet instead of the secondary eclipse of a binary. We reproduce the previous simulation using the same distribution for $\omega$, $i$, and $R_{\star}$ as for binaries. We used the period distribution and eccentricity (within the envelope displayed in Fig. \ref{PerEccPrior}) distribution of giant planets discovered to date by radial velocity (for $m_{p}\sin i > 0.3\,$\Mjup) as provided by the Exoplanet Data Explorer \citep{2011PASP..123..412W} and the radius distribution of \textit{Kepler} giant transiting candidates\footnote{We assumed here that the $\sim$ 19\% of false-positives \citep{2013arXiv1301.0842F} do not significantly bias the radius distribution within the considered range of radii.} \citep[for $6\,\Rearth < r_{p} < 22\,\Rearth$,][]{2012arXiv1202.5852B}. For each simulated giant planet, we computed the impact parameters of both the transit ($b_{tr}$) and occultation ($b_{occ}$). Then, we considered an occulting-only planet if \citep{2010arXiv1001.2010W}:
\begin{eqnarray}
b_{tr} &=& \frac{a}{R_{\star}}\cos(i) \left(\frac{1-e^{2}}{1+e\sin\omega}\right) > 1 + \frac{r_{p}}{R_{\star}}, \label{eq3}\\
b_{occ} &=&\frac{a}{R_{\star}}\cos(i) \left(\frac{1-e^{2}}{1-e\sin\omega}\right) < 1 - \frac{r_{p}}{R_{\star}}, \label{eq4}
\end{eqnarray} where $r_{p}$ is the planetary radius. In the present case, we rejected grazing occultations that are too shallow to reproduce even an Earth-size transit, compared with the binary simulation for which we kept all the grazing eclipse. For each occulting-only giant planet, we computed the occultation depth ($\delta_{occ}$), assuming no thermal emission from the planet \citep{2006ApJ...646.1241R}: 
\begin{equation}
\delta_{occ} = A_{g}\left(\frac{r_{p}}{a_{occ}}\right)^{2},
\end{equation}
where $A_{g}$ is the geometric albedo, supposed to be 0.1 for the majority of close-in giant planets \citep{2011ApJ...729...54C}, and 
\begin{equation}
a_{occ} = a\left(\frac{1-e^{2}}{1-e\sin\omega}\right)
\end{equation}
 is the separation between the star and the planet during the occultation.

\subsection{Results and comparison with other occurrences}

Assuming the occurrence rate of giant planets reported by \citet{2011arXiv1109.2497M}, we found that 0.009 \% $\pm$ 0.002\% of solar-type stars should harbor a giant planet that only presents the occultation, as seen from the Earth. When accounting only for those who present an occultation deeper than 1~ppm, the occurrence decrease to $0.005 \pmÊ0.001$ \%. All the fractions and occurrences of transiting and/or occulting giant planets are listed in Table \ref{table}. As for the binary and triple, uncertainties were estimated by considering the uncertainty of our simulation assuming a Poisson noise and the uncertainty on the occurrence of giant planets from \citet{2011arXiv1109.2497M}. These results can be compared with the expected $235\times(1-19\%) \sim 190 $ giant transiting planets in the \textit{Kepler} field \citep{2012arXiv1202.5852B}, after accounting for 19\% of false positive \citep{2013arXiv1301.0842F}. Assuming 156 000 stars observed by \textit{Kepler}, we can expect from our result 331 $\pm$ 44 giant planets. This discrepancy (at 3.2-$\sigma$) might be explained by the difference in the occurrence rate of planets between the \textit{Kepler} survey \citep{2012ApJS..201...15H, 2012A&A...545A..76S, 2013arXiv1301.0842F} and the radial velocity surveys \citep[e.g.][]{2011arXiv1109.2497M, 2012ApJ...753..160W}. Assuming the occurrence of giant planets from \citet{2013arXiv1301.0842F}, i.e. 5.12\% $\pm$ 0.55\%, we expected \textit{Kepler} to have found 175 $\pm$ 19 giants planets (including $\sim 35$ grazing giant planets), in better agreement with the observed number of candidates.

\subsection{Distributions of occulting-only giant planet}

The distribution of occulting-only giant planets in occultation depth, duration, orbital period, inclination, eccentricity, and argument of periastron are displayed in Fig. \ref{Posterior}. They first reveal that occulting-only giant planets are mimicking sub-Earth objects \citep[like Kepler-37~b,][]{2013Natur.494..452B}, with depths up to a few tens of ppm, with the exception being for a much higher geometric albedo than considered here \citep[e.g.][]{2011A&A...536A..70S}. Then, this type of false positive (even if there are undiluted planets, their characteristics might be misinterpreted) would present a short transit with a median of about 1.7 hours with a median orbital period of about ten days. They should present a moderate eccentricity with a median value of $\sim$0.3.

\section{Discussion and conclusion}
\label{discussion}

We report in this paper a new configuration of false positives involving eclipsing binaries for which only the secondary eclipse occurs. By simulating three secondary-only eclipsing binaries presenting different apparent transit depths, we showed that this false positive can mimic a grazing planetary transit in an eccentric or circular orbit and thus pass unnoticed through a light-curve inspection. We then simulated a population of binary and giant planets and found that $0.061\% \pm 0.017\%$ and 0.009 \% $\pm$ 0.002\% of solar-type stars harbor a secondary-only eclipsing binary or transiting giant planet, respectively. To evaluate the impact of this configuration of false positives in the context of the \textit{Kepler} mission, we simulated secondary-only eclipsing binaries and occulting-only giant planets using the \textit{Kepler}-capability detection model of \citet{2013arXiv1301.0842F}. We find that up to $43.1 \pm 5.6$ KOIs can be mimicked by this configuration of false positives. This corresponds to $1.9 \pm 0.2$\% of the total KOIs identified by \citet{2012arXiv1202.5852B}, thereby re-evaluating the global FPP of the \textit{Kepler} mission from $9.4 \pm 0.9$\% \citep{2013arXiv1301.0842F} to $11.3 \pm 1.1$ \%. These scenarios of false positives do not change the global FPP reported by \citet{2013arXiv1301.0842F} significantly but should be taken into account when validating planet candidates \citep[as in][]{2012ApJ...761....6M}. The detailed numbers of KOIs mimicked by each scenario considered in this paper are listed in Table \ref{table}, and the different apparent sizes of the mimicked KOIs are listed in Table \ref{KOIsize}. These results show that this configuration of false positive scenario tends to mimic giant planets. These false positives scenarios are more likely to occur for the long-period KOIs, as discussed in section \ref{DiscutDistribSOEB}.\\

\begin{table}[]
\caption{Expected radius of KOIs mimicked by secondary-only eclipsing binary or occulting-only giant planets.}
\begin{center}
\begin{tabular}{lc}
\hline
\hline
Expected size of KOI & Number of mimicked KOIs\\
\hline
Earth-size (0.8 -- 1.25 \Rearth) & $0.2 \pm 0.2$\\
Super-earth (1.25 -- 2 \Rearth) & $1.2 \pm 1.0$\\
Small neptunes (2 -- 4 \Rearth) & $5.1 \pm 1.9$\\
Large neptunes (4 -- 6 \Rearth) & $10.5 \pm 2.8$\\
Giant planets (6 -- 22 \Rearth) & $26.0 \pm 4.5$\\
\hline
\hline
\end{tabular}
\end{center}
\label{KOIsize}
\vspace{-0.5cm}
\end{table}

As displayed in Fig. \ref{Posterior} the expected depths of an occulting-only giant planet is less than a few tens of ppm, which only concerns the shallowest part of the \textit{Kepler} candidates. Only $0.4\pm 0.4$ KOI is expected to be mimicked by this scenario. This type of false positive is expected to be more significant in the next-generation space-based transit surveys, like \textit{PLATO} \citep{2012EGUGA..14.7033R}, whose objective is to reach the ppm-level accuracy for most of the targets, observing much brighter stars than \textit{CoRoT} and \textit{Kepler}.\\

We stress that our simulations strongly depend on our current knowledge of the binary population, which is mostly based on the results from \citet{2010ApJS..190....1R}. Even if the authors performed a rigorous characterization of multiplicity of solar-type stars within 25 pc, their results are based on a relatively small statistics involving about 200 binaries and 33 triple systems. Moreover, the stellar multiplicity in transit-survey fields might be different than for the very local neighborhood. A careful statistical analysis of the thousands of binaries \citep[eclipsing or presenting some beaming, ellipsoidal, or reflexion effects;][]{2012ApJ...746..185F} observed in the \textit{Kepler} and \textit{CoRoT} fields will permit an even better understanding of the stellar multiplicity in the galaxy. The ESA \textit{Gaia} mission or LSST survey will also be able to provide a large number of binaries in different regions of the MilkyWay which will greatly strengthen the statistics on binary populations \citep{2012IAUS..282...33E}. This study of stellar multiplicity is also important for improving the priors used for false positives in the planet-validation process.\\

For observed secondary-only eclipsing binaries, the reference epoch of eclipse matches the secondary eclipse. It is thus expected that any radial velocity follow-up will find a significant variation in antiphase with the ephemeris \citep[see discussion about KOI-419 and KOI-698 in][]{2012A&A...545A..76S}. Therefore, this might explain the nine cases of high-amplitude, radial-velocity variation in antiphase that were found in the first fields of \textit{CoRoT} as reported by \citet{2009A&A...506..501C}, \citet{2012A&A...539A..14E}, \citet{2012A&A...538A.112C}, and \citet{2012Ap&SS.337..511C}. Interestingly, four of these candidates present transit-like events with periods longer than or about ten days, as expected, but three present a period of about six days and two others have a period of about two days. If these short-period antiphase candidates are actually secondary-only EB, this would imply that we have underestimated the occurrence of this false-positive scenario in the short-period range. If they are finally more common than we assumed here, these short-period eccentric EB for which only the secondary eclipse is seen should also be present in the ground-based transit surveys such as Super-WASP \citep{2007MNRAS.375..951C,2011PhDT........22T}, HATNet \citep{2007ApJ...656..552B}, and NGTS \citep{2012SPIE.8444E..0EC}.

\begin{acknowledgements}
We thank the anonymous referee for fruitful comments that helped us to substantially improve the quality of this paper. We also thank J. Johnson and T. Morton for their constructive comments about this paper. A.~S., P.~F., and N.~C.~S. acknowledge the support by the European Research Council/European Community under the FP7 through Starting Grant agreement number 239953. NCS also acknowledges the support from Funda\c{c}\~ao para a Ci\^encia e a Tecnologia (FCT) in the form of grant reference PTDC/CTE-AST/098528/2008. RFD is supported by the CNES.

\end{acknowledgements}

\end{document}